\documentclass[fleqn,10pt]{wlscirep}
\usepackage[utf8]{inputenc}
\usepackage[T1]{fontenc}
\title{Calibration of scanning acoustic microscopy for the differentiation between unstable and stable atherosclerotic plaques by X-ray fluorescence imaging}

\author[1,2*]{Peter Modregger}
\author[1,2]{Mallika Khosla}
\author[1,2]{Prerana Chakrabarti}
\author[1]{\"Ozg\"ul \"Ozt\"urk}
\author[3]{Kathryn M. Spiers}
\author[4,5]{Mehmet Burcin Unlu}
\author[4]{B\"ukem Tano\"ren}
\affil[1]{University of Siegen, Physics Department, 57072 Siegen, Germany}
\affil[2]{Deutsches Elektronen-Synchrotron DESY, Center for X-ray and Nano Science CXNS,  22607 Hamburg, Germany}
\affil[3]{Deutsches Elektronen-Synchrotron DESY, 22607 Hamburg, Germany}
\affil[4]{Acibadem University, Department of Natural Sciences, Istanbul, 34684, Turkey}
\affil[5]{Hokkaido University, Faculty of Engineering, North-13 West-8, Kita-ku, Sapporo, Hokkaido, 060-8628, Japan}
\affil[*]{peter.modregger@uni-siegen.de}

\keywords{X-ray fluorescence, scanning acoustic microscopy, atherosclerotic plaques}

\begin{abstract}
Although cardiovascular diseases are the leading cause of death globally, non-invasive and inexpensive diagnostic tools for the identification of associated unstable atherosclerotic plaques are not yet available. Scanning acoustic microscopy offers a high potential to fill this critical gap in patient care. However, convincing validation and calibration of this technique requires high resolution maps of Ca concentrations of atherosclerotic plaques. Here, we demonstrate that synchrotron radiation-based X-ray fluorescence imaging with micrometer spatial resolution can provide such a gold standard.
\end{abstract}
\begin{document}

\flushbottom
\maketitle
\thispagestyle{empty}

\section{Introduction}

According to World Health Organization cardiovascular diseases are the leading cause of death globally~\cite{WHO2018} and unstable atherosclerotic plaques account for 15\% to 20\% of associated heart strokes~\cite{Benjamin2017}. Atherosclerosis, beginning in the early teenage years, develops over the course of 50 years. Adaptive intimal thickening together with accumulation of lipid pools is commonly observed from 20 to 30 years of age in the coronary arteries. A fibrous cap, which is formed over the lipid-rich necrotic cores, may become very thin at a few sites. This lesion, which is called thin-cap fibroatheroma (TCFA), usually is a vulnerable plaque. A TCFA with a fibrous cap thickness of $<65\,\mu$m~\cite{Virmani2000} and with the plaque calcification $<$0.45 of the total volume is considered unstable~\cite{Nandalur2007}. Microcalcifications or spotty calcifications within the plaque are responsible for cardiovascular complications rather than stabilized millimeter-sized or larger calcifications~\cite{Vengrenyuk2006}. Thus, the availability of a diagnostic tool that can distinguish between stable and unstable plaques is crucial. 

However, a clear identification of the plaques via a non-invasive and inexpensive technique is still an unsolved issue. Some conventional imaging methods such as computed tomography and echocardiography can detect advanced calcifications, while others like magnetic resonance imaging, micro-optical coherence tomography or positron emission tomography can identify early calcifications despite certain limitations~
\cite{Shanahan2007,Hjortnaes2013}. These modalities are either expensive or involve ionizing radiation and no modality with sufficient tissue penetration provides both satisfactory morphological and chemical information about tissues on the subcellular level~\cite{Nam2015}. On the other hand, ultrasound imaging with high axial and lateral resolutions of around 20–100~$\mu$m, a good penetration depth of around 5~mm, and low cost, is widely used popular for the observation of soft tissue, however, only morphological information can be provided. Besides, the signal detection capability of conventional ultrasonography has to be increased for the detection of microcalcifications, since high echo signals from such small surfaces are not available~\cite{Kamiyama2008}.

Scanning acoustic microscopy (SAM) utilizes focused ultrasound signals to measure material specific speed of sound or acoustic impedance for 2D imaging of tissue morphology and mechanical properties on a micrometer scale. It has been demonstrated that SAM can distinguish collagen-rich from calcified regions in atherosclerotic plaques~\cite{Bilen2018}, which is key for the identification of potentially unstable plaques. Since SAM constitutes a non-invasive, non-ionizing and inexpensive method it offers a high potential for translation into diagnostic tool. However, SAM delivers information that is only proportional to Ca concentrations. Proper calibration for translating SAM into a quantitative method has not been performed due to a lack of a gold standard. 

Here, we propose high resolution X-ray fluorescence (XRF) imaging as method providing quantitative elemental concentrations to meet this demand. The basic principle of XRF is as follows. An X-ray photon incident on a sample can excite an inner-shell electron leaving behind an unoccupied state. This state can then be filled by an outer-shell electron, which releases a fluorescence X-ray photon upon transition. The energy of that fluorescence photon is characteristic of the element. Since the number of fluorescence photons is proportional to the elemental content in the sample, XRF can be used to quantify the corresponding elemental concentrations~\cite{Beckhoff2006}.

\section{Methods}

\subsection{Plaque samples}
Five (N=5) human carotid atherosclerotic plaques samples were received from patients undergoing carotid endarterectomy with informed consent under the ethical approval of the Istanbul University Medical Faculty (No:2018/952). Samples were fixated in 2\% formaldehyde, which is known to change acoustical properties by less than 2\%~\cite{Miura2015}. Samples were cut into segments of approximately (2x1x1)~mm$^3$ or smaller.

\subsection{SAM imaging}

\begin{figure}[htbp]
\centering
\includegraphics[width=0.45\textwidth]{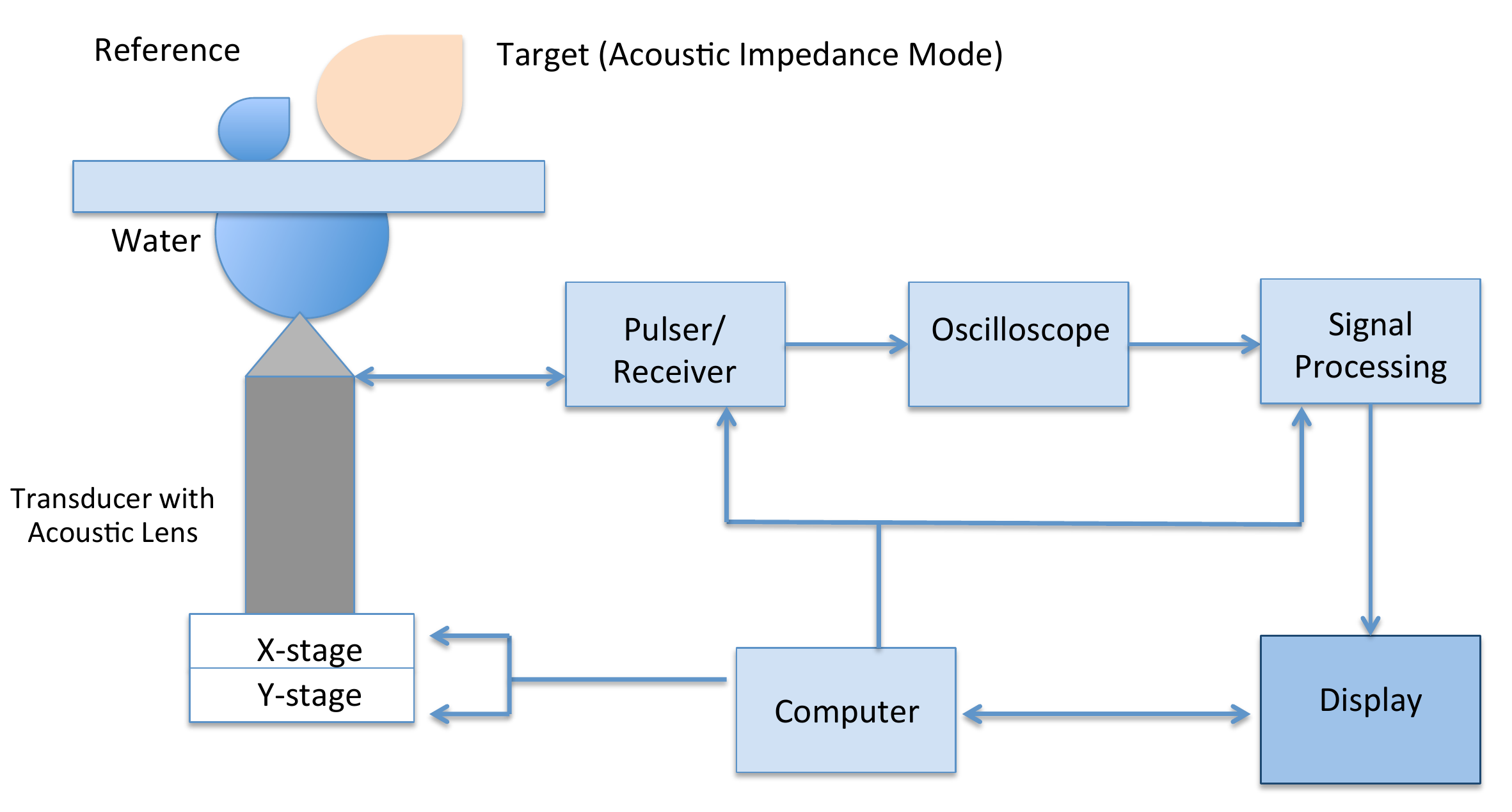}
\caption{Sketch of the setup for scanning acoustic microscopy.}
\label{fig:sam_sketch}
\end{figure}


An AMS-50SI scanning acoustic microscope (Honda Electronic, Toyohashi, Japan) in acoustic impedance (AI) mode was used to acquire the SAM data. An 80 MHz transducer with a piezo crystal provided both the generation and the detection of ultrasound signals, which were focused by coupled sapphire lens. The spot size was 17~$\mu$m and the focal length was 1.5~mm. The penetration depth was approximately 20~$\mu$m, the generated signal was –6 dB and the expected signal-to-noise ratio (SNR) was around 25 dB. Distilled water was used to couple the sapphire lens with the substrate. An X-Y stage was used for two-dimensional scanning of the transducer. The signals, reflected from the target's surface and also the reference's surface, were analyzed by an oscilloscope. Finally, the acoustic impedance maps with a lateral resolution of approximately 20~$\mu$m were obtained with 300 x 300 sampling points for a maximum area of 4.8~mm x 4.8~mm in less than 2~minutes (see Fig.~\ref{fig:sam_xrf}a). More details about the principle of SAM in AI mode can be found in~\cite{Bilen2018}.

\subsection{XRF imaging}

\begin{figure}[htbp]
\centering
\includegraphics[width=0.45\textwidth]{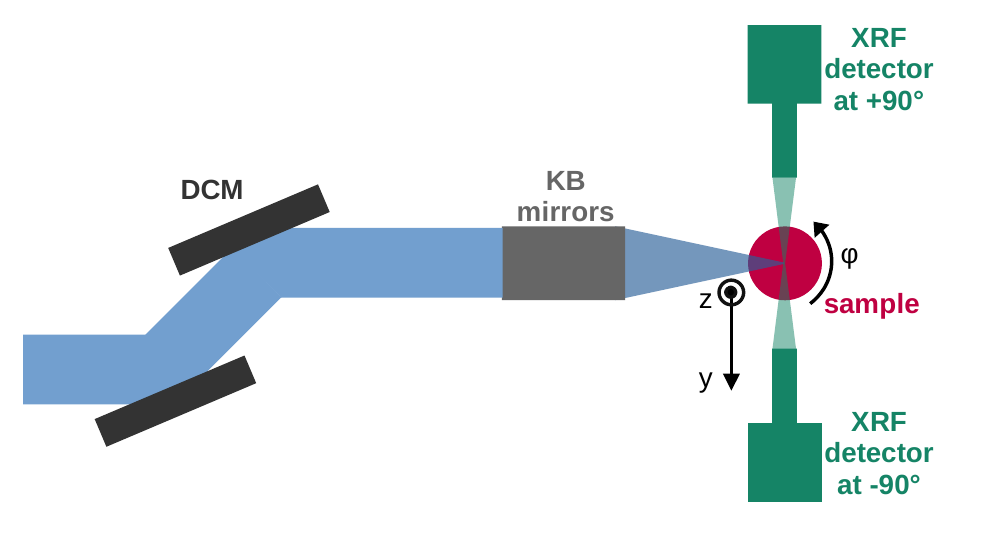}
\caption{Sketch of the setup for X-ray fluorescence imaging.}
\label{fig:xrf_sketch}
\end{figure}

XRF measurements were performed at the microhutch endstation of the P06 beamline (Fig.~\ref{fig:xrf_sketch}) at Petra III in Hamburg, Germany~\cite{Falkenberg2020}. A double crystal monochromator (DCM) was used to select an incident photon energy of 12~keV, which was just above the K absorption edge of As (11.9~keV). Two orthogonal Kirkpatrick-Baez (KB) mirrors focused the X-ray beam to approximately (1x1)~$\mu\mathrm{m}^2$ at the position of the sample. Detection of the fluorescence signal was provided by two silicon drift detectors (SII Vortex EM, Hitachi High-Tech Science Corporation, Tokyo, Japan) connected to digital pulse processors (Xspress3, Quantum Detectors, Harwell Oxford, UK) positioned at a distance of about 6~mm from the sample and at $\pm 90$\textdegree{} from the incident beam to minimize the contribution of scattering.

Samples were placed in Kapton tubes with 0.5~mm to 1~mm diameter and mounted on a sample manipulation stage, which allowed for horizontal\/ vertical (i.e., $y$ and $z$) stepping as well as rotation (i.e., $\phi$). The samples were scanned through the focal X-ray spot in steps of 5~$\mu$m over an area of (1 x 1.6)~mm$^2$ with a dwell time of 10~ms resulting in a total scan time of approximately 10~min per projection. This was repeated 3 times for $\phi = 0$\textdegree{}, 45\textdegree{} and 90\textdegree{}, which -- in combination with the 2 opposite XRF detectors -- resulted in 6 independent projections per sample.

\begin{figure}[htbp]
\centering
\includegraphics[width=0.45\textwidth]{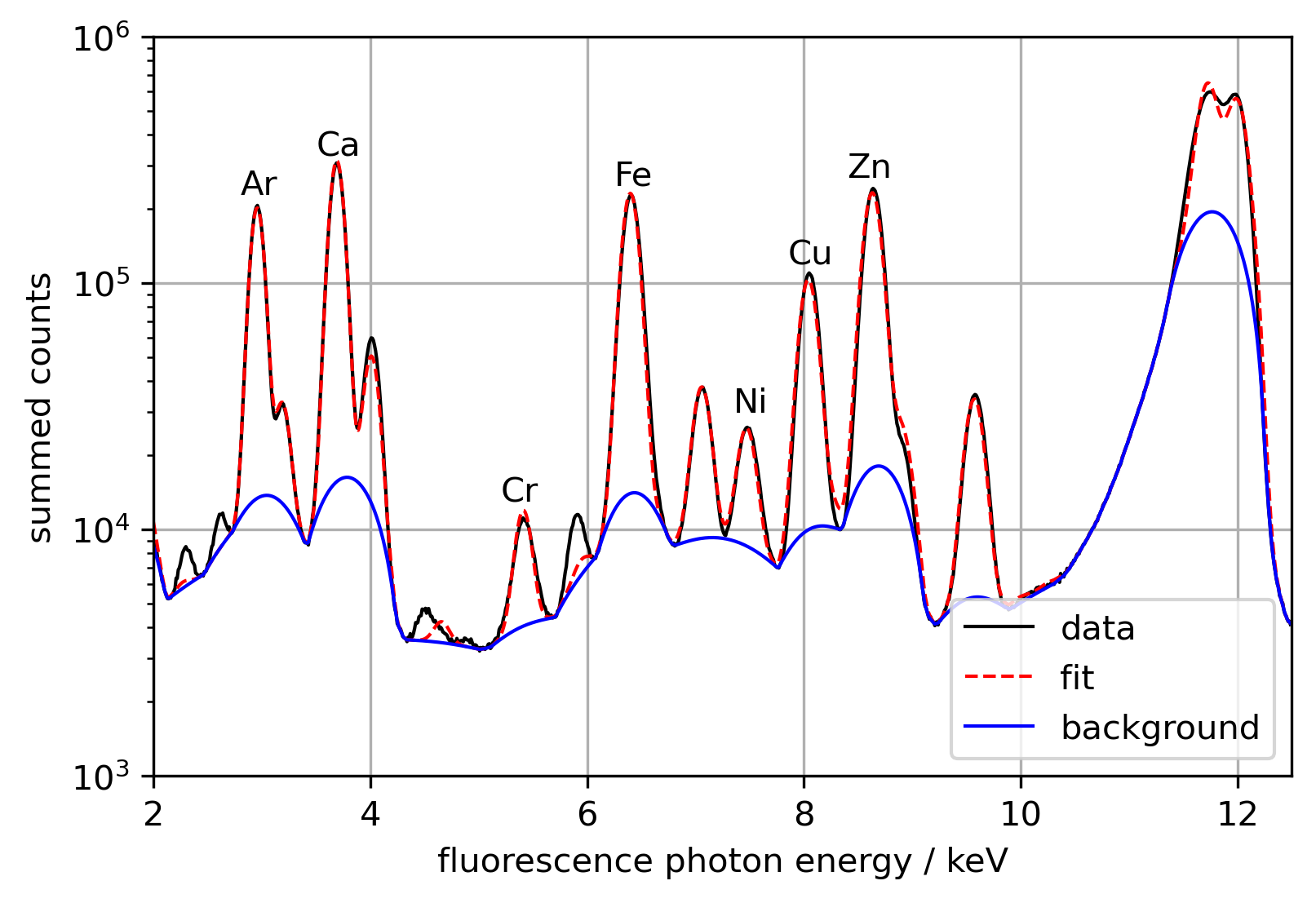}
\caption{Semi-logarithmic plot of a XRF spectrum summed over all scan points of the atherosclerotic plaque sample shown in Fig.~\ref{fig:sam_xrf} and the fitted curve. Labeled peaks correspond to K$_{\alpha}$-lines and unlabeled peaks to K$_{\beta}$-lines of the adjacent element. The strength of background subtraction was adjusted in a way to reproduce the relative counts of the Ca K$_{\alpha}$ and K$_{\beta}$-lines.}
\label{fig:xrf_fit}
\end{figure}

XRF data was analyzed with PyMca~\cite{Sole2007}, which included the determination of fundamental X-ray beam parameters (i.e., incident photon flux and sample to detector distance) with a CaF$_2$ calibration foil (Micromatter Technology, Inc.) of known thickness and concentration (areal density = $20.1\,\frac{\mu\mathrm{g}}{\mathrm{cm}^2}$). The XRF data of the plaques was then fitted to retrieve the quantitative elemental concentrations of Ca and relative elemental concentrations of Cr, Fe, Ni, Cu and Zn. Fig.~\ref{fig:xrf_fit} shows the XRF spectrum of a plaque summed over the entire field of view and the corresponding fit.

\section{Results}

\begin{figure}[htbp]
\centering
\includegraphics[width=0.4\textwidth]{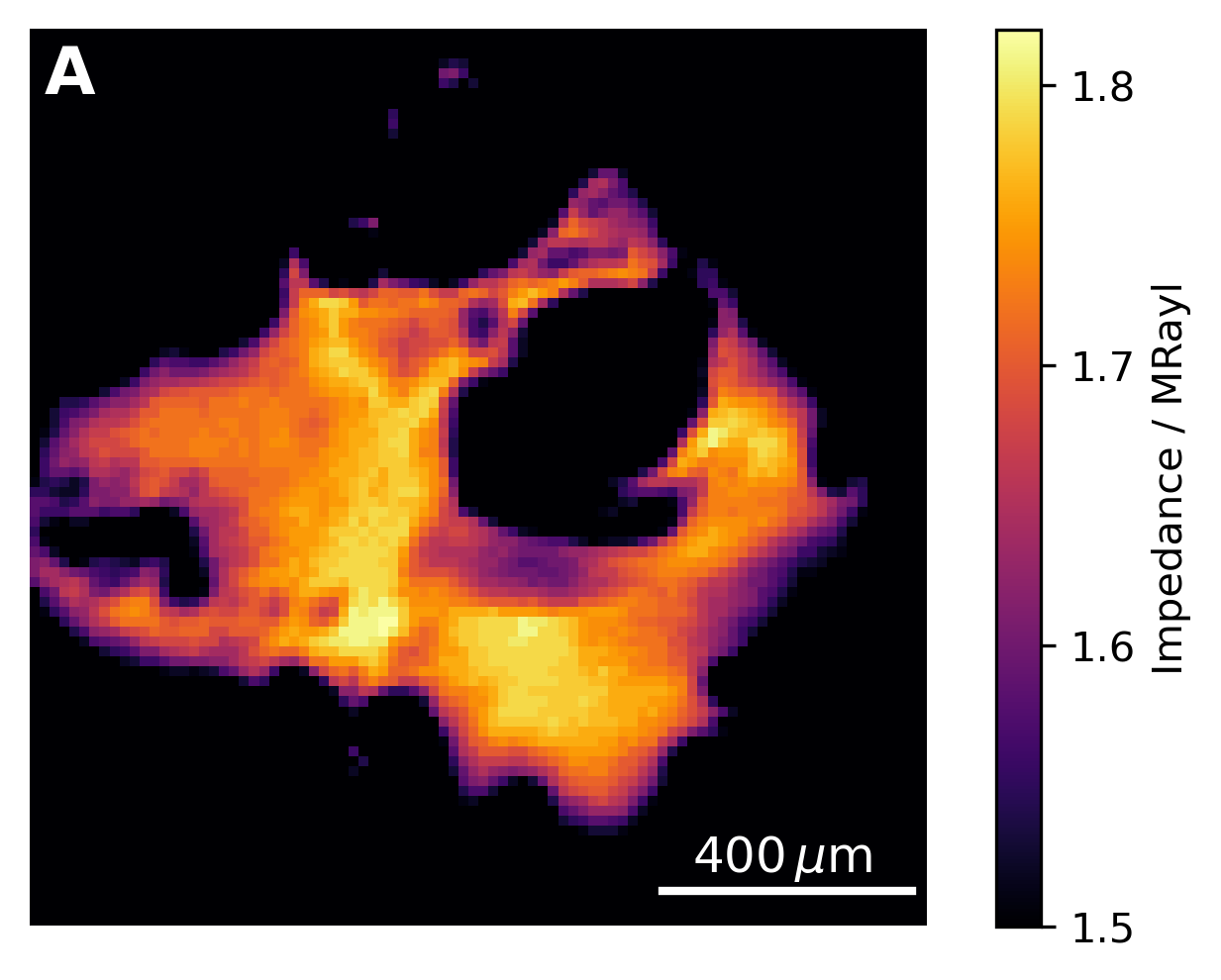}
\includegraphics[width=0.43\textwidth]{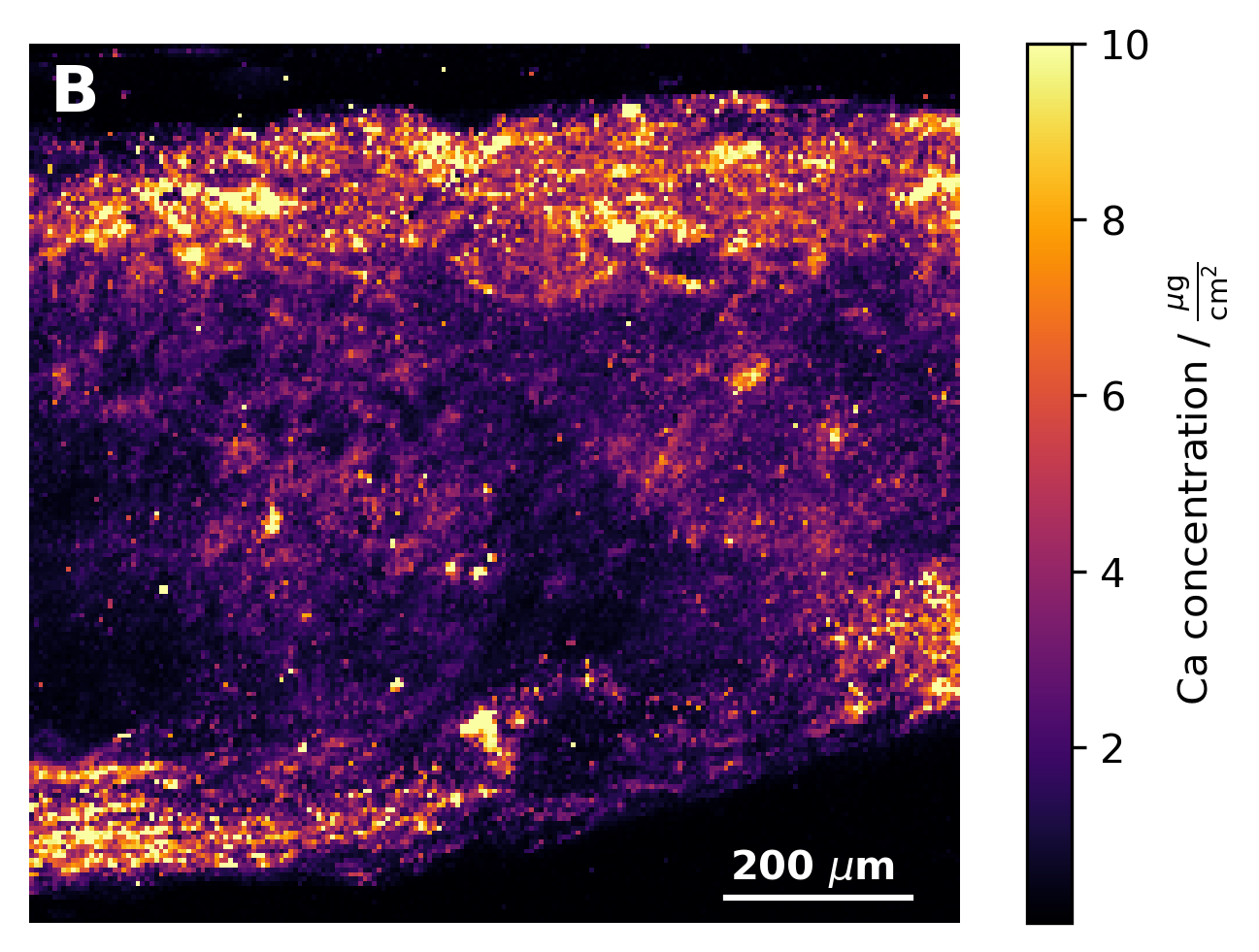}
\caption{Images of a human atherosclerotic plaque. (a) Acoustic impedance as determined by SAM shown with a lower threshold of 1.5~mRayl, the impedance of water. (b) Ca concentration as determined by XRF.}
\label{fig:sam_xrf}
\end{figure}

\begin{figure}[htbp]
\centering
\includegraphics[width=0.43\textwidth]{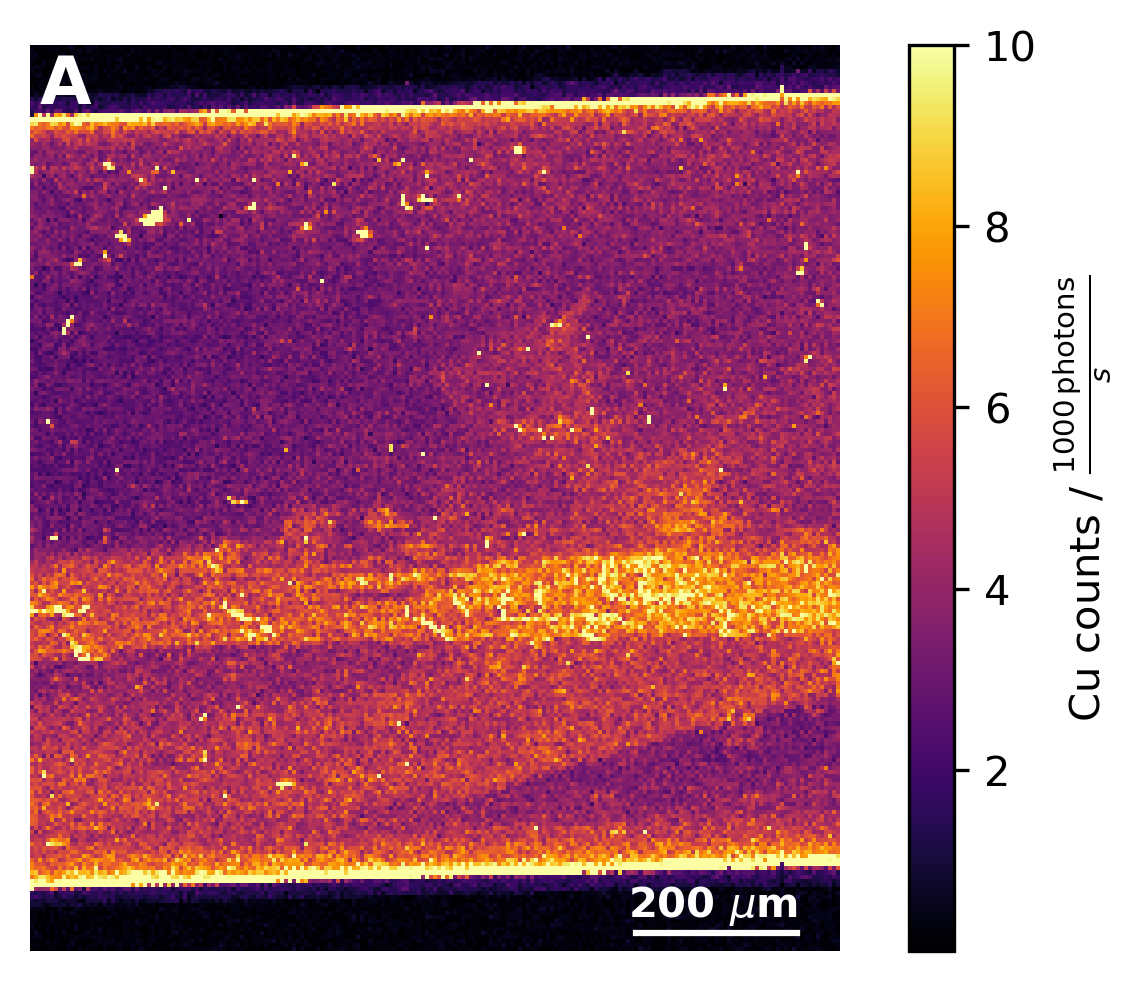}
\includegraphics[width=0.4\textwidth]{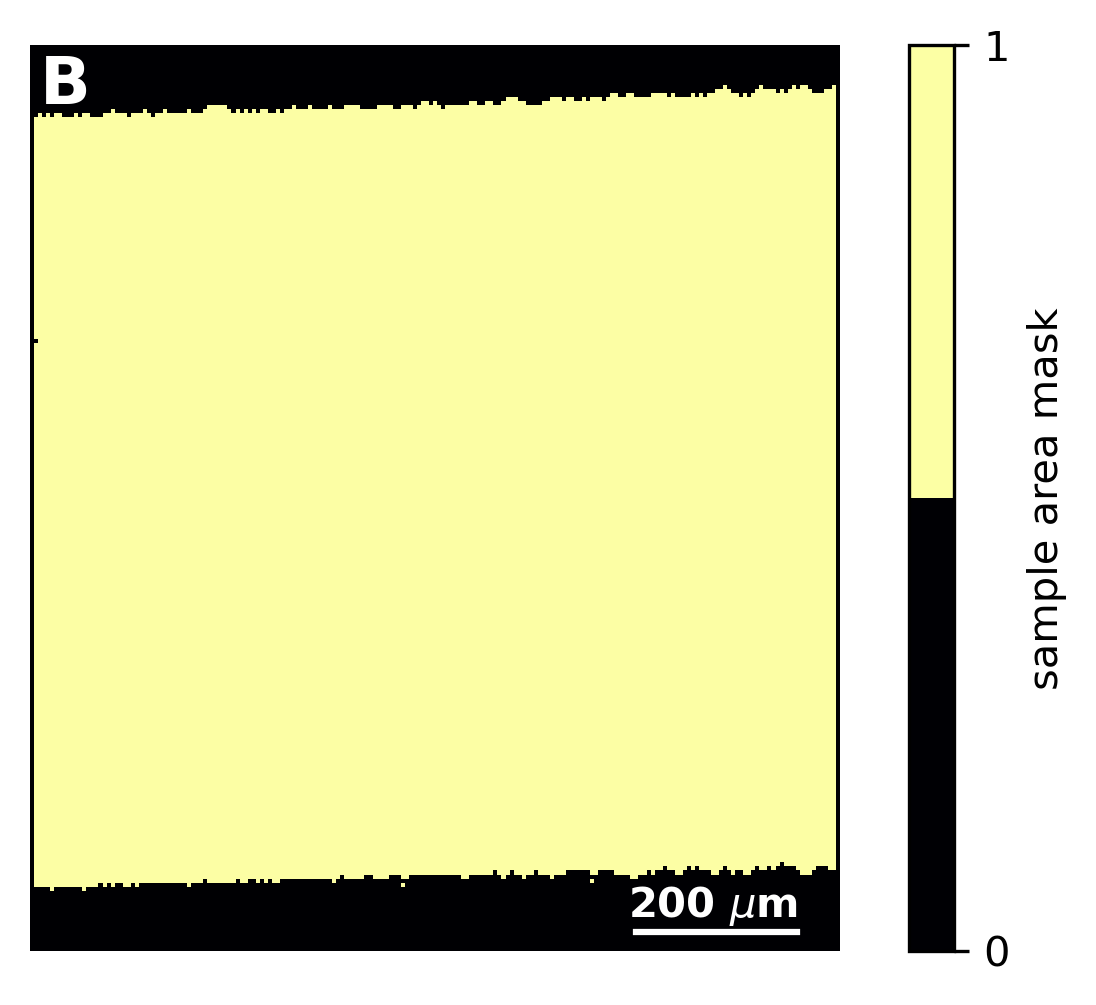} 
\caption{Determination of the sample area in XRF. (a) Cu counts of the same sample shown in Fig.~\ref{fig:sam_xrf}b. (b) Binarization of the image in (a) using a threshold of 2000~counts.}
\label{fig:xrf_cu}
\end{figure}

In order to demonstrate the possibility of using XRF data for the calibration of SAM we have calculated the correlation between median impedance values (Fig.~\ref{fig:sam_xrf}a) and median Ca concentrations (Fig.~\ref{fig:sam_xrf}b) for the 5 human atherosclerotic plaque samples. To this end the sample region for SAM data was segmented using a threshold of 1.5~MRayl, which is the impedance value of water. Further, the same goal was achieved for XRF data by utilizing the fairly homogeneous distribution of Cu in the samples (Fig.~\ref{fig:xrf_cu}a), which allowed for binarization with a threshold of 2000~photons/s (Fig.~\ref{fig:xrf_cu}b).

\begin{figure}[htbp]
\centering
\includegraphics[width=0.45\textwidth]{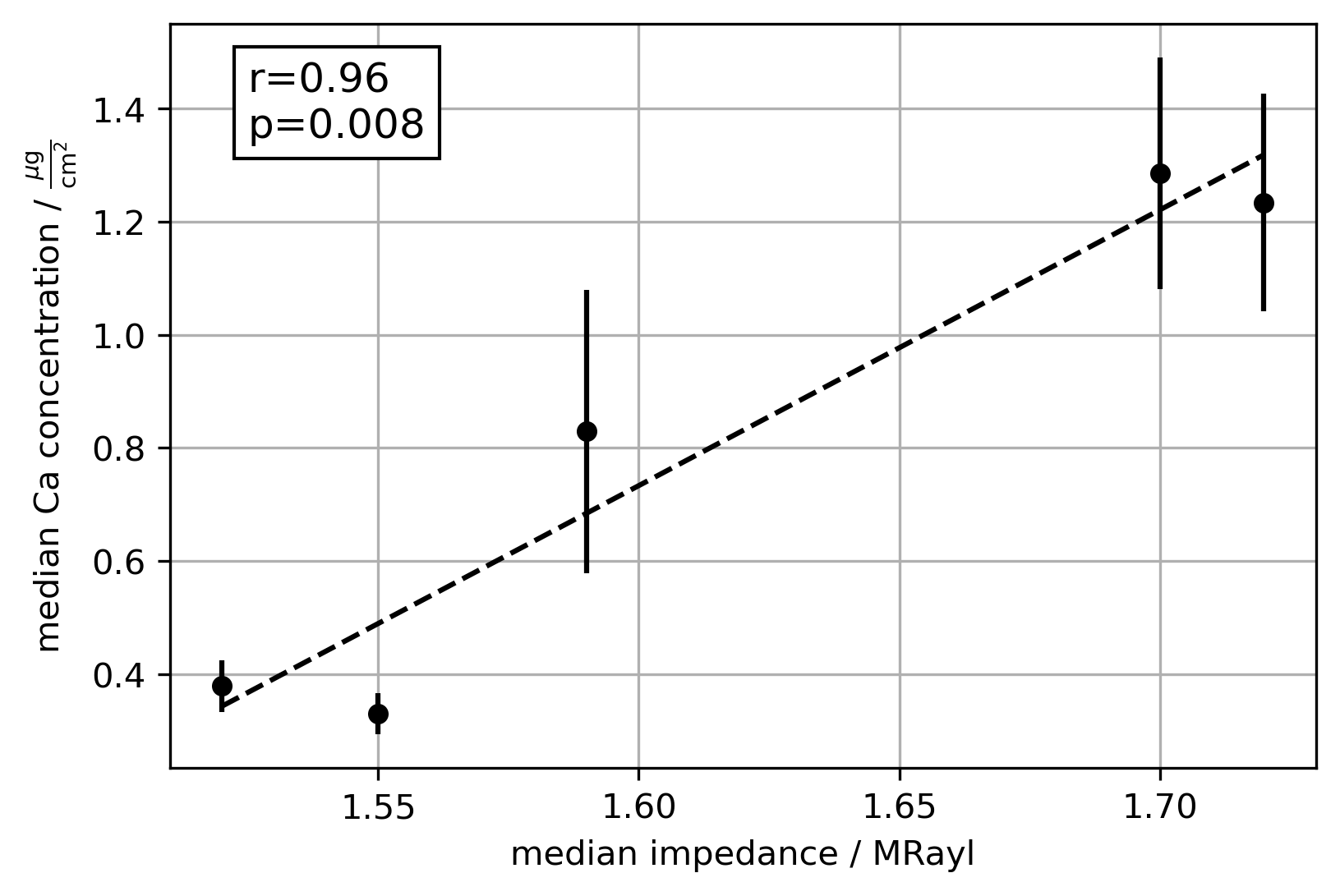}
\caption{Correlation between the median impedance determined by SAM and the median Ca concentration determined by XRF. This curve represents a model for the calibration of SAM by XRF.}
\label{fig:corr}
\end{figure}

Figure~\ref{fig:corr} shows an excellent correlation (r=0.96) with high  statistical significance (p=0.008) between median impedance and median Ca concentrations. Each of the 6 XRF projections per sample were treated as independent measurements of the Ca concentrations, which allowed us to determine the uncertainties in the Ca concentrations. The strong correlation represents a conceivable calibration curve for SAM impedance values and, thus, validates the presented approach.

\section{Discussion}

\begin{figure}[htbp]
\centering
\includegraphics[width=0.43\textwidth]{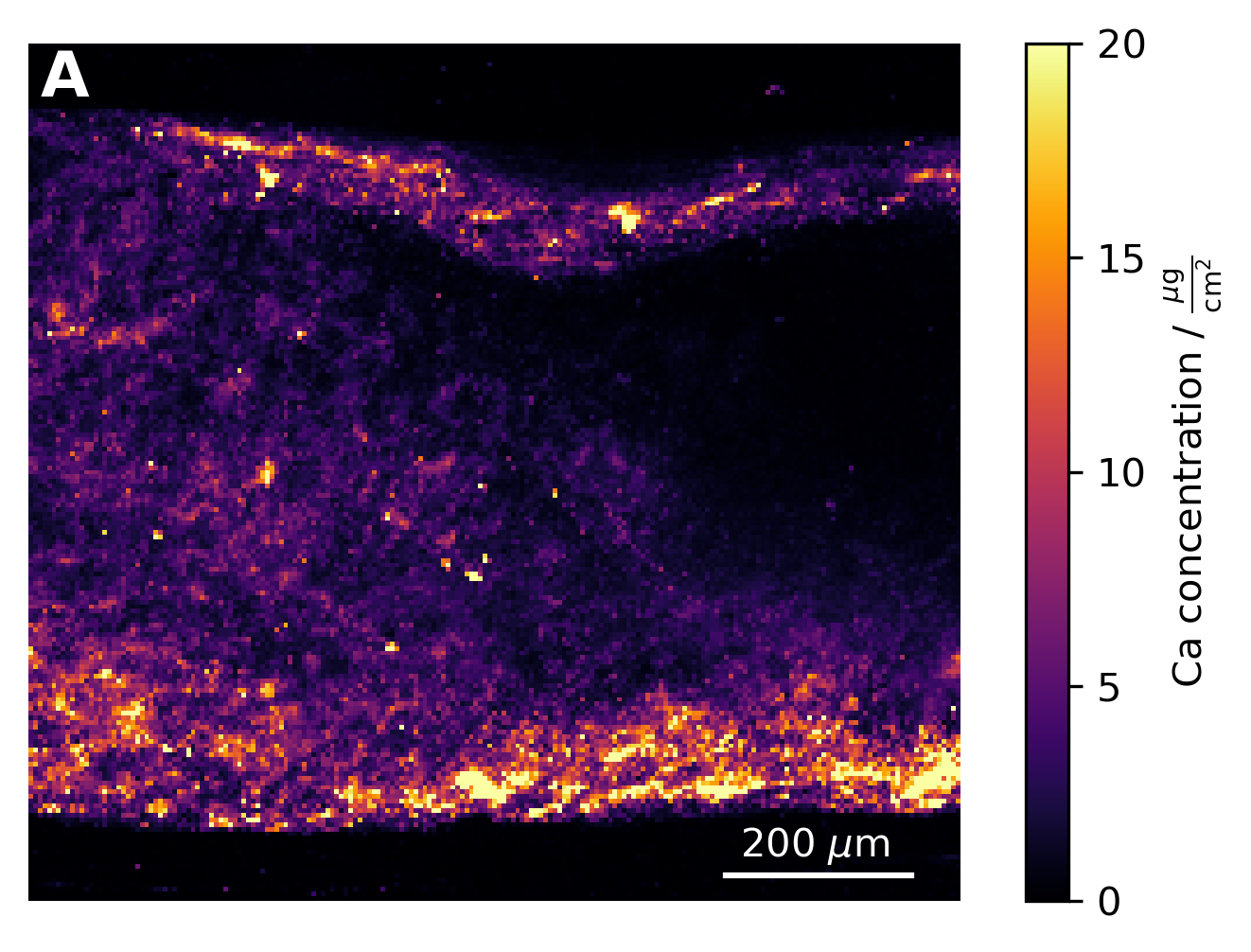} 
\includegraphics[width=0.43\textwidth]{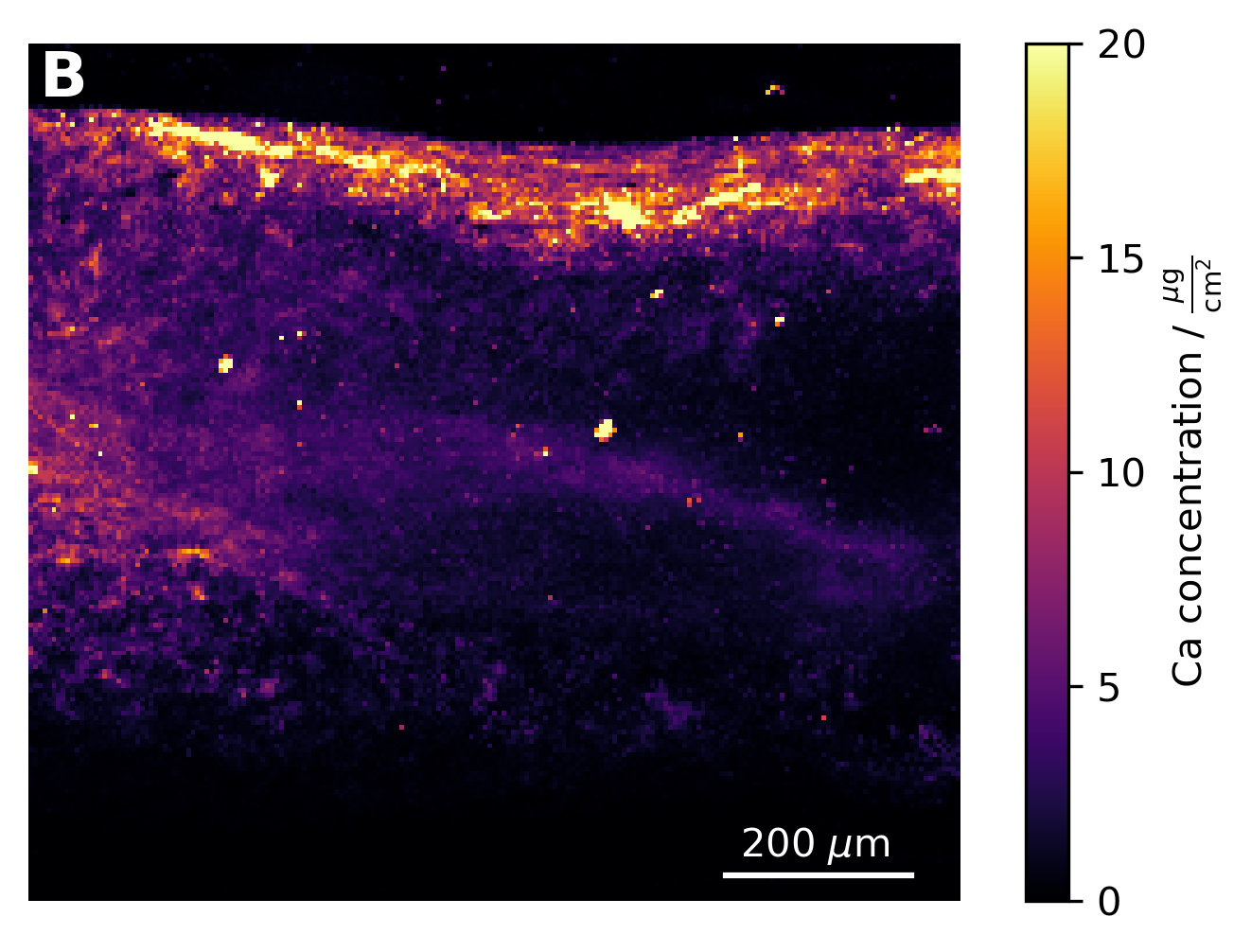} 
\caption{Impact of self-absorption to the XRF signal of atherosclerotic plaques. Self-absorption appears as a difference between the Ca concentration maps measured at $\phi =$ -90\textdegree{} (a) and $\phi=$ +90\textdegree{} (b).}
\label{fig:xrf_sa}
\end{figure}

In XRF imaging, self-absorption relates to the fact that emitted fluorescence photons can be absorbed by the sample itself. This leads to a reduction of measured fluorescence photons and, consequently, to an underestimation of the concentration of the corresponding element. As Ca has a comparatively low fluorescence energy for the K$_{\alpha 1}$-line of 3.692~keV, a noticeable impact of self-absorption was expected. Figure~\ref{fig:xrf_sa} exemplifies this impact on two Ca concentrations maps imaged at opposite sides of the sample. Here, self-absorption led to increased uncertainties for atherosclerotic plaques with a median Ca concentration above 0.8~$\frac{\mu\mathrm{g}}{\mathrm{cm}^2}$, which is reflected by the corresponding larger error bars in Fig.~\ref{fig:corr}. However, the correlation between SAM and XRF was still robust. 

Nevertheless, atherosclerotic plaques samples with relatively small Ca concentrations were chosen for this study in order to mitigate the impact of self-absorption that was unknown beforehand. To assess the degree of calcification of the utilized samples we estimated the areal Ca concentration of a fully calcified plaque with a thickness of 50~$\mu$m, a mass density of $1.45\,\frac{\mathrm{g}}{\mathrm{cm}^3}$ and a Ca weight percentage of 0.115~\cite{Rahdert1999} as approximately 800~$\frac{\mu\mathrm{g}}{\mathrm{cm}^2}$. In the present study the highest detected areal Ca concentrations in clearly defined micro-calcifications were about 60~$\frac{\mu\mathrm{g}}{\mathrm{cm}^2}$. Thus, the samples with the highest degree of calcification were about 8\% calcified, which is a factor of 5 smaller than the 45\% demarcation between between stable and unstable plaques~\cite{Nandalur2007}. 

A further limitation of this study related to the limited number of samples (N=5), which was in line with the proof of concept approach laid out here.

\section{Conclusions \& Outlook}

We have demonstrated that synchrotron radiation-based XRF imaging with micrometer spatial resolution can provide a quantification standard for SAM measurements of human atherosclerotic plaques. This was established by the robust correlation between the quantitative Ca concentrations determined by XRF and the relative acoustic impedance determined by SAM. Thus, this study constitutes a vital step towards establishing SAM as a non-invasive and inexpensive diagnostic tool for the identification of unstable atherosclerotic plaques.

Future studies will include a larger number of plaque samples with a wider range of calcification degrees to increase the statistical power and range of the calibration. Further, this study showed that the impact of self-absorption in XRF for the determination of Ca concentrations was manageable. This will allow for the inclusion of both stable and unstable plaque samples.

\section*{Funding}
Scanning acoustic microscopy studies were supported by a grant from the Directorate of Presidential Strategy and Budget of Turkey (Project Number: 2009K120520).

\section*{Conflicts of interest}
There are no conflicts to declare.

\section*{Author Contributions}
Peter Modregger: conceptualization, formal analysis, investigation, supervision, visualization, writing – original draft. Mallika Khosla: formal analysis, investigation, writing – review \& editing. Prerana Chakrabarti: investigation, writing – review \& editing. \"Ozg\"ul \"Ozt\"urk: conceptualization, resources, writing – review \& editing. Kathryn M. Spiers: investigation, writing – review \& editing. Mehmet Burcin Unlu: funding acquisition, writing – review \& editing. Bukem Bilen: conceptualization, formal analysis, resources, writing – review \& editing.

\section*{Acknowledgements}
We acknowledge DESY (Hamburg, Germany), a member of the Helmholtz Association HGF, for the provision of experimental facilities. Parts of this research were carried out at PETRA III and we would like to thank K.~V.~Falch, J.~Garrevoet and G.~Falkenberg for assistance in using the P06 beamline. Beamtime was allocated for proposal I-20200444. We further like to thank Christina Ossig, Dennis Br\"uckner and Michael Stuckelberger (all DESY) for their input on quantitative XRF analysis.

\bibliography{xrf} 

\end{document}